\documentclass[aps,prl,twocolumn,superscriptaddress,showpacs,amsmath,amssymb,longbibliography]{revtex4-2}
\usepackage[mathlines]{lineno}
\usepackage{amsthm}
\usepackage{amsmath}
\usepackage{amssymb}
\usepackage{graphicx}
\usepackage{bm}
\usepackage{color}
\usepackage[dvipsnames]{xcolor}
\usepackage{enumitem}
\usepackage{mathrsfs}
\usepackage{verbatim}
\usepackage{bbold}
\usepackage{braket}
\usepackage{lipsum}
\usepackage[colorlinks, breaklinks, 
linkcolor=OrangeRed,
citecolor=RoyalBlue,
urlcolor=RoyalBlue]{hyperref}
\usepackage{ulem}

\usepackage{color}

\usepackage{tikz}
\usepackage{natbib}
\usepackage{graphicx}

\newcommand\redsout{\bgroup\markoverwith{\textcolor{brown}{\rule[0.6ex]{4pt}{1.0pt}}}\ULon}
\begin{document}
\title{Einstein relation for  subdiffusive relaxation in Stark chains}

\author{P. Prelov\v{s}ek}
\affiliation{Jo\v zef Stefan Institute, SI-1000 Ljubljana, Slovenia}
\author{S. Nandy}
\affiliation{Jo\v zef Stefan Institute, SI-1000 Ljubljana, Slovenia}
\author{M. Mierzejewski}
\affiliation{Institute of Theoretical Physics, Faculty of Fundamental Problems of Technology, Wroc\l aw University of Science and Technology, 50-370 Wroc\l aw, Poland}

\date{\today}
\begin{abstract}

We investigate  chains of interacting spinless fermions subject to a finite external field $F$ (also called Stark chains) and focus on the regime where
the charge thermalization follows the subdiffusive hydrodynamics. First, we study reduced models conserving 
the dipole moment and derive an explicit Einstein relation which links the subdiffusive transport coefficient  with the correlations
of the dipolar current. This relation explains why the decay rate, $\Gamma_q$, of the density modulation with wave-vector $q$  shows $q^4$-dependence.  In the case of  the Stark model, a
similar Einstein relation is also derived and tested using various numerical methods. They
confirm an exponential reduction of the transport coefficient with increasing $F$. 
On the other hand, our study of the Stark model indicates that upon increasing $q$  there is a crossover from subdiffusive behavior, $\Gamma_q \propto q^4$, to  the normal diffusive relaxation,  $\Gamma_q \propto q^2$,
at the wave vector $q^*$ which vanishes for  $F \to 0$.  
\end{abstract}
\maketitle
{\it Introduction.}  
Macroscopic systems driven by finite external forces/fields are usually described within the extended concept of 
local equilibrium via local thermodynamic 
parameters as, e.g.,  the
temperature and  chemical potential.  Such a description can fail when we are dealing with  systems isolated 
from the environment.  Prominent recent  examples are those of the Stark systems of interacting fermions 
(which is fermion systems in the presence of a finite field),
as realized in tilted cold-atom lattices \cite{guardado20,scherg21,kohlert23}.  Such systems reveal several novel 
theoretical insights and challenges,  discussed mostly within one-dimensional (1D) models. Since noninteracting particles subject to a finite external field, $F$, exhibit Stark localization, the problem shares some similarities with the many-body localization  (MBL) (for an overview see \cite{nandkishore15,alet18,abanin19,sierant24}) and it is known as the Stark MBL  \cite{schulz19,nieuwenburg19,taylor20,nehra20,zhang21,morong21,wang21,
doggen21,gunawardana22,zisling22,wei22}. It is also well established that effective models, derived at large $F$
 from the Stark model (usually by invoking the Schrieffer–Wolff transformations), can exhibit Hilbert space fragmentation \cite{sala20,khemani20,yang20,moudgalya22,moudgalya221,skinner22} that violates the eigenstate thermalization hypothesis (ETH) \cite{deutsch91,srednicki94,dalessio16}.
In systems with strongly fragmented Hilbert space,  the latter can be linked with the emergence of additional integrals of motion \cite{rakovszky20, lydzba24}.

Still at moderate $F$, the cold-atom experiments \cite{guardado20} as well as numerical simulations of
1D models \cite{nandy23} reveal the hydrodynamic relaxation of the inhomogeneous particle distributions towards the steady state,
which corresponds to the infinite-temperature ($ T \to \infty$) equilibrium. For small wave-vectors $q \ll 1$ the relaxation rates follow
a particular subdiffusive (SD) law $\Gamma_q  \propto q^4$, rather than the 
normal diffusive behavior,  $\Gamma_q \propto q^2$. This is well consistent with the fact that at $F \ne 0$, the  dipole moment emerges in macroscopic systems as an additional conserved quantity \cite{sala20,feldmeier20,scherg21,kohlert23,nandy23} and the phenomenological 
description can be given in terms of the fracton hydrodynamics \cite{nandkishore19,feldmeier20,gromov20,
burchards22,guo22,glodkowski23}.  In spite of a wide consensus that such a description is appropriate
for  isolated driven systems, so far the theoretical studies are mostly  based on phenomenological 
hydrodynamic approaches. There are so far very few quantitative  results for the subdiffusion in 
microscopic models \cite{nandy23} as well as theoretical attempts to express 
the  SD transport coefficient $D_S$ \cite{guardado20,zechmann22} in terms of the response functions.

In this Letter, we present the analysis, focused on the particle/density relaxation and anomalous diffusion in the hydrodynamic regime.
We consider  the standard 1D Stark model,  i.e., the chain of interacting
spinless fermions subject to a finite external field $F$,  as well as the  effective models which involve extended  
pair-hopping (EPH) interactions and strictly conserve the dipole moment, $P$.
 On the one hand, from numerical results for the dynamical density structure factor 
$S_q(\omega)$, we extract the relaxation rates $\Gamma_q$  of the density modulation. These rates reveal subdiffusive transport 
for $ q \to 0$ as well as  the value of the corresponding SD coefficient $D_{S}$. On the other hand, employing the
memory-function (MF) formalism, we derive the Einstein relation that expresses  $\Gamma_{q \to 0}$
in terms of the uniform ($q=0$) correlations of the normal current, $J_N$, and the dipolar current, $J_D$. 
If $P$ is  conserved the response function is determined  solely by $J_D$ and the relaxation follows the 
SD relation   $\Gamma_q = D_S q^4$.
In the EPH model,  $J_D$ is a translationally-invariant operator that governs the relaxation. 
In the case of the full Stark model, $J_D$ dominates the response for large $L$ (for $F >0$) when one observes  the emergent conservation of the dipole moment.
The Einstein relation as well as the numerical results for $D_{S}$ in both models are  tested with 
alternative numerical approaches.  It should be emphasized that the derived
Einstein relations remain valid beyond the considered models and even beyond 1D, which we mostly discuss below.  
Moreover, our numerical results in Stark chains for larger $q >0 $ reveal the crossover from SD  $\Gamma_q \sim D_S q^4$ to normal diffusion $\Gamma_q \sim D_N q^2$ at $ q \sim q^{*}(F)$ with vanishing $q^{*}(F \to 0)$,  consistent with some phenomenological theories \cite{guardado20,zechmann22}.

 {\it Dynamical density-modulation  relaxation.}  
In the following, we  study two 1D lattice models of interacting spinless fermions, as they 
emerge in the presence of the finite external field $F$, whereby the chain has $L$ sites and 
open boundary  conditions (OBC).  The coupling to the field enters the Hamiltonian via  $H'=F P$ where $P$ is 
the dipole moment, $P = \sum_l (l-L/2)  n_l$, and  $n_l$  is the particle number operator at site $l$. 

Isolated macroscopic Stark  systems at finite $F > 0$ 
develop (heat up)  towards  a homogeneous steady state $\langle n_l \rangle =\bar n = N/L$
 ($N$ representing the total particle number), corresponding to  $T \to \infty$ equilibrium. 
 Further on, we analyze dynamics  of the periodic density modulation 
$n_q = \sum_l e^{iql} \tilde n_l/ \sqrt{L}$, $\tilde n_l =n_l - \bar n$, and related correlation function $\phi_q(\omega)$,
\begin{eqnarray}
\phi_q(\omega) &=& \frac{\chi_q(\omega) - \chi^0_q}{ \omega} =  \frac{ - \chi^0_q}{\omega +M_q(\omega)}, 
\nonumber \\
\chi_q(\omega) &=&   \frac{i}{\beta} \int_0^\infty dt e^{i\omega t} \langle [n_{-q}(t),n_q] \rangle, \label{phiqo}
\end{eqnarray}
whereby we define dynamical susceptibilities $\chi_q(\omega)$ and \mbox{$\chi^0_q = \chi_q(\omega =0)$ } 
that remain nonzero even at $\beta =1/T \to 0$. 
In this limit, $\phi_q(\omega)$ is related with  the standard dynamical structure
factor $S_q(\omega) = \mathrm{Im} \phi_q(\omega)/\pi $. In general, $\phi_q(\omega)$ can be 
represented in terms of the memory function (MF), $M_q(\omega)$, that determines the profile relaxation rate
$\Gamma_q  =  \mathrm{Im}M_q(\omega = 0)$, which we later on extract also from numerical results for 
$\phi_q(\omega)$. 

{\it Einstein relation.}
An analytical step towards $M_q(\omega)$ can be made using 
the MF formalism \cite{mori65,forster75,gotze72,jung07}, with the introduction of the general 
scalar product of two operators, $(A|B)$, and the Liouville operator ${\cal L} A =[H,A]$. 
In the case of $\beta \to 0$, the latter scalar product reduces to thermodynamic average, i.e., 
$(A|B) \sim \langle A^\dagger B \rangle$. Within this formalism one can express  the correlation function as
$\phi_q(\omega) =(n_q |({\cal L} -\omega)^{-1} | n_q)$.  The memory function
 can be written in the hydrodynamic regime $q \to 0$ \cite{forster75} (in analogy to the perturbation theory \cite{gotze72}) 
 as
 \begin{equation}
M_q(\omega) = ( {\cal L} n_q| ({\cal L}  - \omega)^{-1} | {\cal L} n_q)/ \chi^0_q. \label{mf}
\end{equation}
Expanding $n_q$ in powers of $q$ one obtains, 
\begin{equation}
{\cal L} n_q \simeq \frac{1}{\sqrt{L}} \left( iq {\cal L}P+iq^2  \frac{i}{2} {\cal L} Q \right), \quad Q=  \sum_l l^2 \tilde n_l, \label{expa}
\end{equation} 
where we assumed conservation of the particle number, \mbox{$ {\cal L} N =0$} with  $N= \sum_l n_l$. The first
term in Eq.(\ref{expa}) represents the normal (uniform) current, \mbox{$J_N= i {\cal L} P$}. It determines the hydrodynamic relaxation ($q\to 0$) in generic systems 
that do not conserve the dipole moment, ${\cal L} P \ne 0$. Namely, one obtains from Eq.(\ref{mf}) the standard Einstein relation \cite{kadanoff63,forster75,bonca95,steinigeweg09}, \mbox{$ \Gamma_q=  q^2   \mathrm{Im} \phi_N(0)/ \chi^0_0=D_N q^2 $},
which links the diffusion constant, $D_N$, with the current correlation function  $\phi_N(\omega)=(  J_N | ({\cal L}  - \omega)^{-1} |  J_N)/L$.  

If the dipole moment is conserved,
${\cal L} P = 0$, then the hydrodynamic relaxation is determined by the second term in the expansion in Eq. (\ref{expa}), which can be interpreted as the dipolar current $J_D= \frac{i}{2} {\cal L} Q$.  
 Similarly, one then obtains then the Einstein relation  \mbox{$ \Gamma_q =D_{S}q^4 $} with $D_S = \mathrm{Im} \phi_D(0)/ \chi^0_0$, however, involving the dipolar currents \mbox{$\phi_D(\omega)=(  J_D | ({\cal L}  - \omega)^{-1} |  J_D)/L$}. To conclude this part, we note that the MF formalism straightforwardly explains  the origin of SD
 in systems with dipole-moment conservation, $ \Gamma_q= D_S q^4$. Similarly to the phenomenological fracton hydrodynamics \cite{nandkishore19,feldmeier20,gromov20,
burchards22,guo22,glodkowski23}, the only assumption we made is the conservation of the dipole moment. 
    
 {\it Extended pair-hopping (EPH) model.} As a first example we  analyze the model where $P$ is strictly a conserved.
Starting from the full Stark models, one can derive at $F \gtrsim 1$ the EPH model,
either via the Schrieffer-Wolff transformation \cite{sala20,khemani20,feldmeier20}, or
expanding the interaction  in the Stark basis \cite{lydzba24},
\begin{equation}
H_{EPH}= \sum_l \zeta_{dr} [c_{l-r}^\dagger c_l c_{l+d+r}^\dagger c_{l+d}  + \mathrm{H.c.}] + H_d. \label{hph}
\end{equation}
Here,  $c^\dagger_l,c_l$ refer to localized Stark states and $H_d$ represents the Hartree-Fock 
diagonal term, while   $\zeta_{dr}$ can be derived for given $F$, at least to lowest order in the interaction 
\cite{lydzba24}.  Here, by construction  ${\cal L} P=0$. One can derive explicit expression for the dipolar current  \mbox{$J_D = \frac{i}{2}
{\cal L}Q =\sum_{dr} \zeta_{dr} 
J_D(d,r)$}, where
\begin{equation}
J_D(d,r)= - r (r + d ) \sum_j ( i c_{j-r}^\dagger c_j c_{j+d+r}^\dagger c_{j+d} + \mathrm{H.c.}).
\label{jdr} 
\end{equation}
It is remarkable that explicit $l$-dependence cancels out and $J_D$ emerges as a translationally-invariant operator.

\begin{figure}[!tb]
\includegraphics[width=0.8\columnwidth]{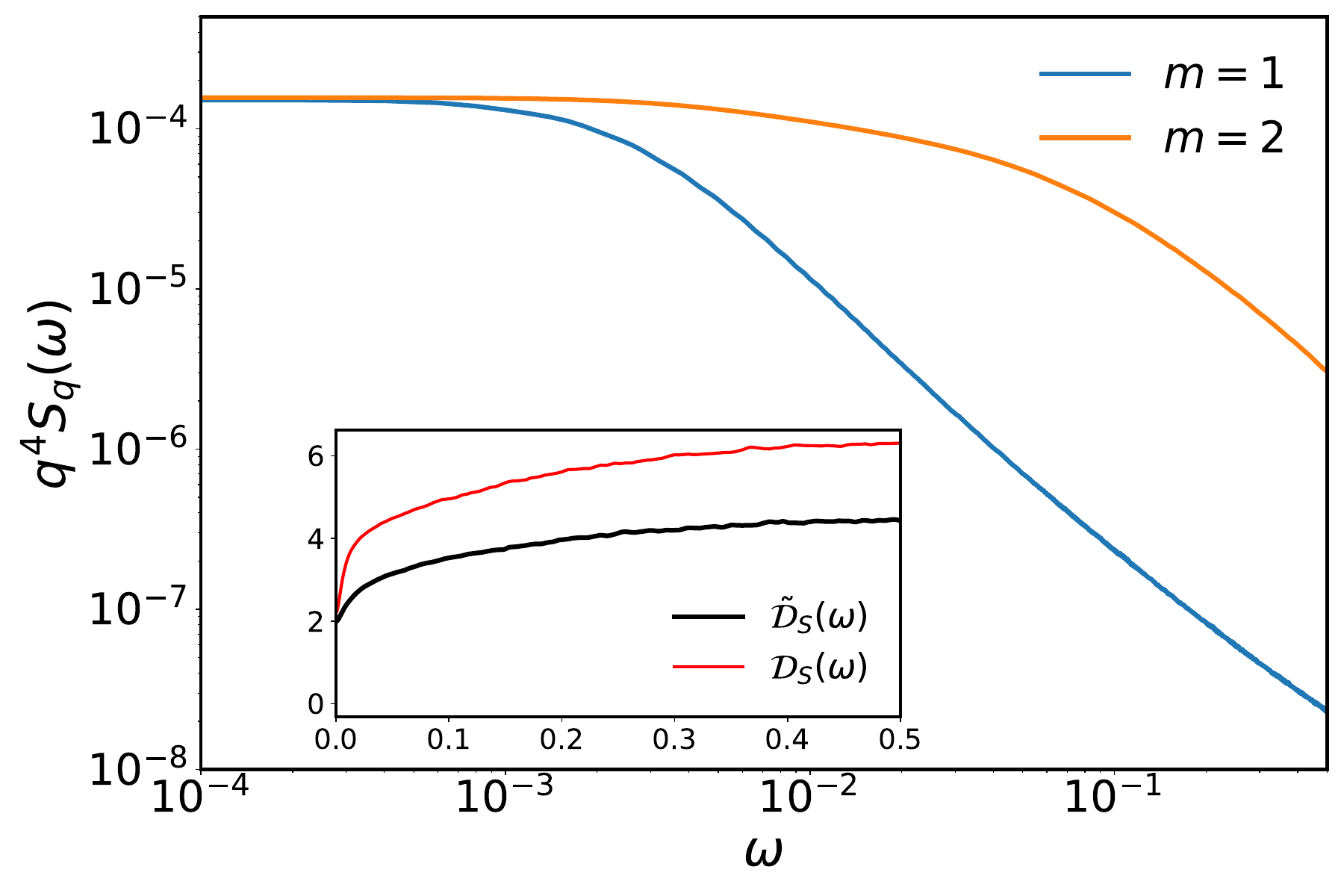}
\caption{Dynamical structure factor $q^4 S_q(\omega) $ for the EPH model with $L=32$ sites, obtained via MCLM  
for the lowest $q = 2 m\pi/L, m=1,2$. The inset shows the extracted $\tilde D_{S} (\omega) = \mathrm{Im}M_{q_1}(\omega) /q_1^4$,
compared to the Einstein-relation result $D_{S}(\omega)=  \mathrm{Im} \phi_D(\omega)/ \chi^0_0$. }
\label{fig1}
\end{figure}

Here, we do not aim to investigate closer the EPH models with realistic parameters, 
but rather test the existence of SD and the direct expression for the coefficient $D_{S}$ 
for a simplified case with $H_d=0$ and (rather arbitrary) assuming $r=1$ and values
$\zeta_{d1} = 1, 0.75, 0.5, 0.25$ for $d=1,...,4$. The motivation for including longer-range $d > 1$ terms is that the 
basic pair-hopping model with only $\zeta_{11}$ is known to exhibit strong Hilbert space fragmentation \cite{sala20,khemani20,moudgalya22,moudgalya221} which invalidates basic ETH  concept,
while additional terms with $\zeta_{d>1,r>1}$ are expected to supress this effect \cite{lydzba24}.
In the following we calculate numerically $\phi_q(\omega)$, Eq. (\ref{phiqo}), using the microcanonical Lanczos method 
(MCLM) for finite $L$ systems \cite{long03,prelovsek11,herbrych12}, employing large number  of Lanczos steps 
up to $ N_L \sim 10^5$,   to achieve the frequency  resolution $\delta \omega \lesssim 10^{-4}$
 which allows for reliable extraction of $M_q(\omega)$ \cite{herbrych12} even for small $q \ll 1$.  
The advantage of the EPH model (relative to the Stark model) is that  in addition to
$N$ we use also conservation of $P$ (choosing  $P=0$) to reduce the Hilbert
space and to reach $L=32$ with $N_{st} \sim 10^7$ basis states. 

In Fig.~\ref{fig1} we present results for the density structure factor $q^4 S_q(\omega)$ 
as calculated via MCLM for two lowest  $q=q_m = 2 m \pi/L, m=1,2$.  Since we consider 
the half-filled system $\bar n=1/2$ with  effective $T \to \infty$, we know analytically 
$\chi^0_q \sim \chi_0^0 \sim \bar n (1-\bar n) = 1/4$.
Results confirm very sharp  peak at $\omega \sim 0$,
being consistent with SD hydrodynamics, i.e., $\pi S_q(\omega \sim 0) \sim \chi_q^0/\Gamma_q 
\sim \chi_0^0/(D_S q^4)$. Moreover, we extract also the corresponding (dynamical) SD coefficient
 $\tilde D_{S} (\omega) = \mathrm{Im}M_q(\omega) /q^4$ for smallest $q=q_1$ and present results in the inset, 
 together with  numerically evaluated Einstein relation $D_S(\omega) = \mathrm{Im} \phi_D(\omega)/\chi^0_0$,
using $J_D$ from Eq.~(\ref{jdr}).  
The agreement is reasonable given that both numerical approaches
can suffer from finite-size effects. Moreover, the considered EPH 
model can still exhibit some features of the Hilbert-space fragmentation \cite{sala20,khemani20,moudgalya22,moudgalya221}, 
which  could influence the presumed ETH.

{\it  Stark model.} 
We turn further to the properties of the prototype Stark model, i.e.,  1D chain of interacting spinless 
fermions in the presence of a finite external field $F$,
\begin{eqnarray}
H &=& t \sum_{i} ( c^\dagger_{l+1} c_l +  c^\dagger_l c_{l+1} ) + V \sum_l \tilde n_{l+1}  \tilde n_{l}  
+ \nonumber \\
&+& V' \sum_l \tilde n_{l+2}  \tilde n_{l}  + F P,  \label{stark}
\end{eqnarray}          
with $\tilde n_l = n_l - \bar n, n_l = c^\dagger_l c_l$. Fermions interact via  the nearest-neighbor
  ($V$) and next-nearest-neighbor  ($V'$) repulsion.
We consider half-filling, i.e.,  $\bar n = N/L  =1/2$, and set $t=1$ as the unit of energy.
We introduce $V' \ne 0$ in order to suppress the integrability (and dissipationless transport)
at $F \to 0$, although the latter effect appears not to be important for $F \gg 0$. 
In the main text, we restrict the numerical results to the case $V=V'=1$, while in \cite{supmat}
we discuss also  results for $V=2, V'=0$. 

In the case of the Stark model, Eq.~(\ref{stark}),
we cannot  apply the same analysis as for EPH model,
since ${\cal L} P \ne 0$ and the conservation of  $P$ emerges only in the thermodynamic limit,
 $L \to \infty$ \cite{nandy23}. Still, one can derive from Hamiltonian.~(\ref{stark}) both contributions
  to $ {\cal L}  n_q$ in Eq. (\ref{expa}):  \mbox{ $J_N= i {\cal L} P=\sum_l J_l  $} and 
  $J_D= \frac{i}{2} {\cal L} Q=J_N/2+\sum_l   lJ_l  $, where $J_l= it c^\dagger_{l} c_{l+1} + {\rm H.c.} $.
 Neglecting possible offdiagonal 
correlations, we can then express the corresponding Einstein relation from Eq.~(\ref{mf}), 
\begin{equation}
M_q(\omega) \simeq [q^2 \phi_N(\omega)+ q^4 \phi_D (\omega)]/\chi^0_0. \label{mqs}
\end{equation}
Since in general $\phi_N(\omega) \ne 0$, we can proceed by showing that for finite $F >0 $
and $L \to \infty$ it follows \mbox{$\phi_N(\omega \to 0)=0 $},
related to emergent conservation of $P$ \cite{nandy23}, as verified also numerically
in \cite{supmat}. Consequently, we expect the at $F>0$ the hydrodynamic $q \to 0$ behavior
will be dominated by SD with $\Gamma_q =  D_S q^4$ with $D_S=  {\rm Im} \phi_D (\omega \to 0)/\chi_0^0$.

\begin{figure}[!tb]
\includegraphics[width=\columnwidth]{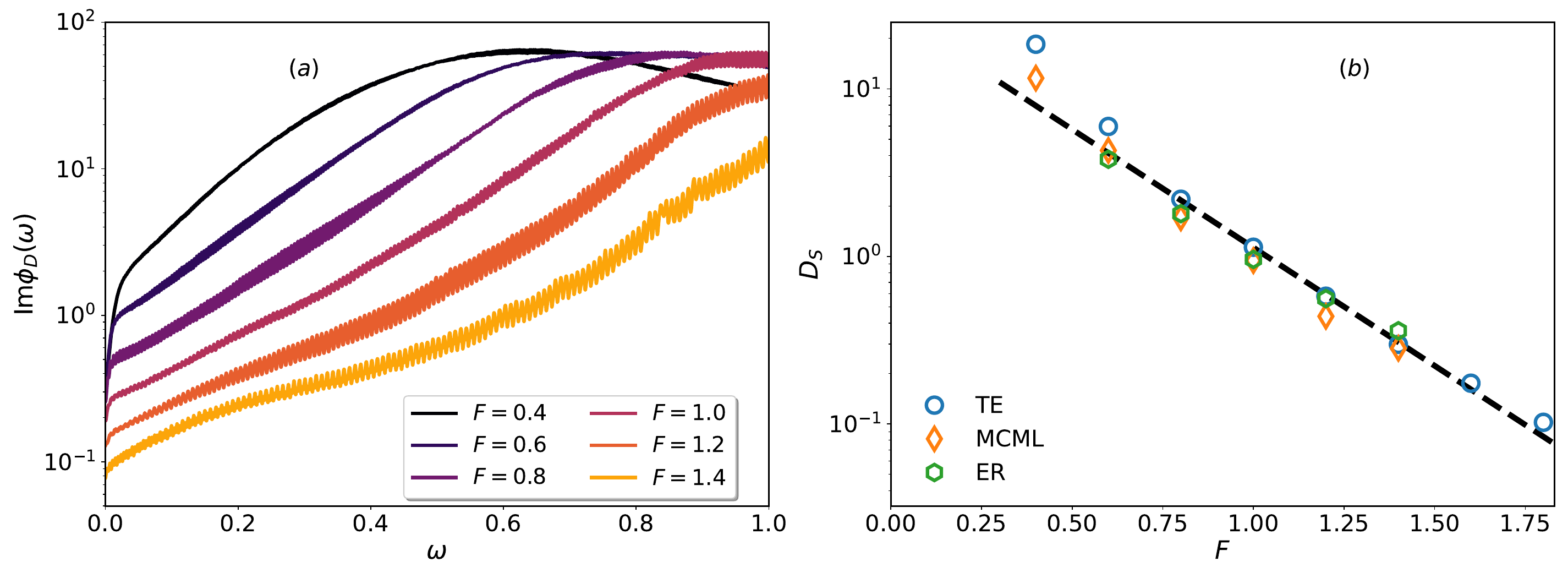}
\caption{(a) Dipolar-current correlations $\mathrm{Im}\phi_D(\omega)$ (in log scale)
as calculated with MLCM for the Stark model on \mbox{$L=28$} sites for different fields $F$. (b) Subdiffusion coefficient $D_S(F)$, calculated  
from the rates $\Gamma_{q_1}$, obtained via:  time-evolution (TE) of the density profile and via  
MCLM on $L=24$ chain,  and directly from the Einstein relation (ER), i.e., from $\phi_D(\omega)$. }
\label{fig2}
\end{figure}

We further present numerical results for $\mathrm{Im} \phi_D(\omega)$  in Fig.~\ref{fig2}(a). 
Since by construction $J_D$  requires OBC,  one cannot guarantee 
$L$-independent  result for \mbox{$\phi_D(\omega \to 0)$}. Still, results in Fig.~\ref{fig2}(a), obtained with MCLM 
on $L=28$ chain, indicate that beyond $F> F_*(L) \gtrsim 0.4$,   there is a well defined 
value $D_S = \mathrm{Im} \phi_D(\omega \to 0)/  \chi^0_0$, revealing already 
an exponential-like dependence on $F$. 
We also note  that $\mathrm{Im} \phi_D(\omega)$  for larger $\omega$, as in Fig.~\ref{fig2}(a), does not 
have a direct relation to $M_q(\omega)$,  since it neglects $\phi_N(\omega)$  
contribution in Eq.~(\ref{mqs}).

Results for the transport coefficient $D_S(F)$ are summarized in Fig.~\ref{fig2}(b).
Besides the results from the Einstein relation evaluated  via $\phi_D(\omega \to 0)$  (see Fig.~\ref{fig2}(a)),
we include also results for $D_S  = \Gamma_{q}/q^4$ obtained from  two alternative approaches 
applied  for $L=24$ chain and the smallest  $q_1= 2 \pi/L$.
Namely, we extract $ \Gamma_{q}$ directly from the MCLM results for
$S(q,\omega \to 0)$,  as well as from the decay of the inhomogeneous density profile
where $F$ is introduced via the time-dependent flux \cite{mierzejewski10,nandy23}. The latter method
evaluates the relaxation rate $\Gamma_{q}$  (see \cite{nandy24} for the details), 
with the advantage of periodic boundary conditions and consequently resolving also very small $D_S$, i.e.,
reaching larger $F\simeq 2$.  Results in Fig.~\ref{fig2}(b) are quantitatively consistent in the
broad regime of $F > F_*(L) \sim 0.4$, confirming the validity of the Einstein relation for SD transport as well as  the exponential dependence of $D_S(F)$.  In \cite{supmat} we present results also for other parameters, in particular for $V=2, V'=0$, where we employ also the boundary-driven open systems \cite{nandy23,nandy24}, allowing
for the analysis of considerably larger $L \leq 50$ as well as the direct evaluation of $D_S$ via the form of the 
nonequilibrium stationary density profile \cite{supmat}. 

\begin{figure}[!tb]
\includegraphics[width=1.0\columnwidth]{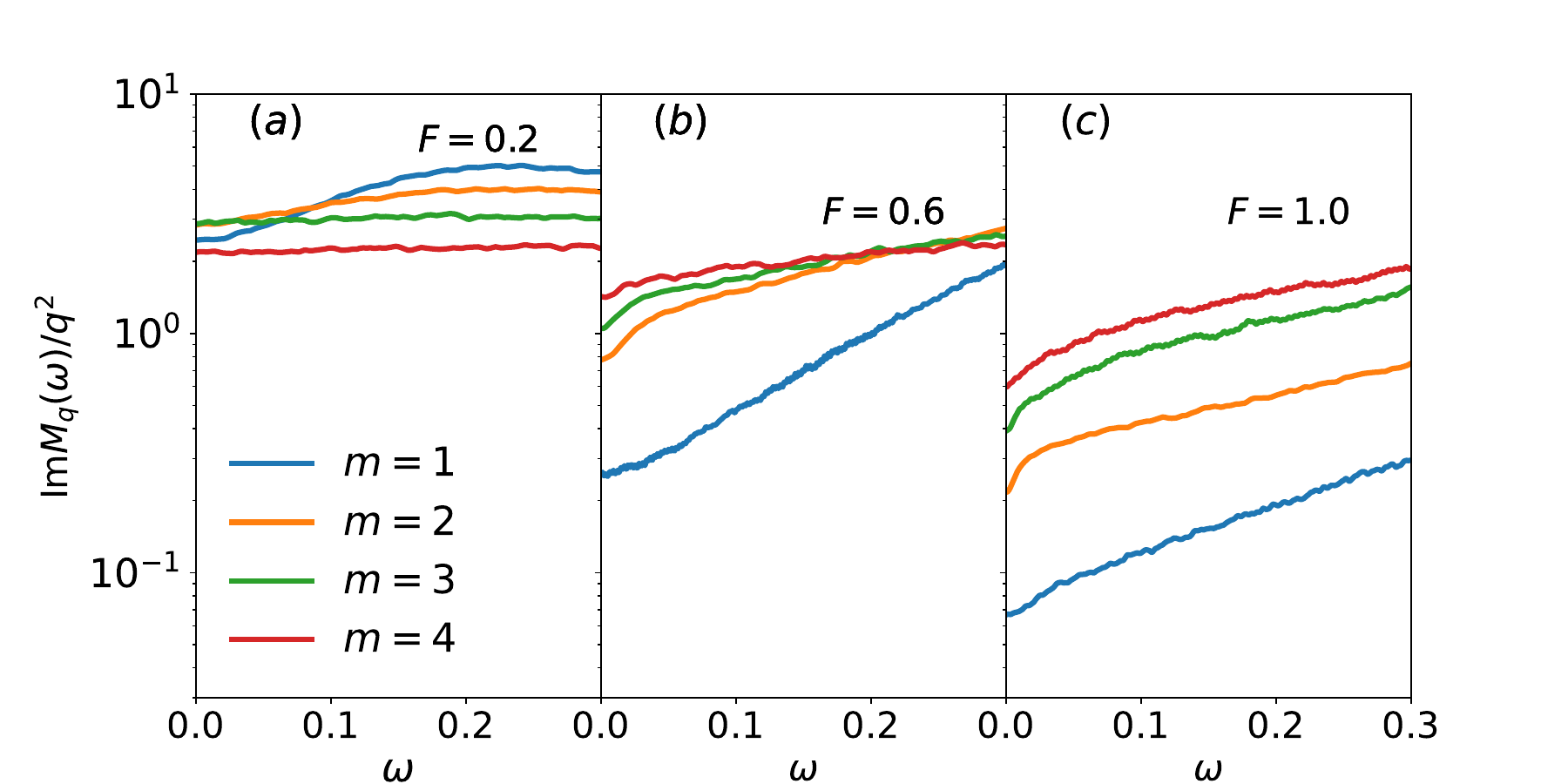}
\caption{ Dynamical relaxation rates $\Gamma_q(\omega)/q^2$ (in log scale),  as extracted from MCLM results 
for $\phi_q(\omega)$ on $L=24$ chain and $q_m =2m \pi/L, m=1 -4$, for three different regimes of $F$.}
\label{fig4}
\end{figure}

{\it Crossover to normal diffusion.}  Let us finally examine closer $M_q(\omega)$ 
at larger $q$  as extracted numerically from $\phi_q(\omega)$, again for 
 parameters $V = V' = 1$. Fig.~\ref{fig4} shows MCLM results for  ${\rm Im} M_q(\omega)/q^2$   
obtained for various $q = 2 m\pi /L$ with  $1\le m\le 4$  and $L=24$ so that $q \leq \pi/3$.  
One can resolve three regimes.  Despite finite-size limitations, results at  small $F = 0.2  <F_*(L) $,  shown in 
    Fig.~\ref{fig4}(a), are approximately consistent with  normal diffusion for all presented $q_m$, 
i.e.,  ${\rm Im} M_q(\omega) \propto q^2$. 
At intermediate $F_*(L) < F < F_c \sim 1$, as in Fig.~\ref{fig4}(b), only
the smallest $q=q_1$ evidently deviates, the latter being the signature of the SD transport 
${\rm Im} M_q(\omega \sim 0) \propto q^4$. Still,  the relaxation functions ${\rm Im} M_q(\omega)/q^2 $
nearly overlap for larger $q$ which can be interpreted as an effective  normal diffusion ${\rm Im} M_q(\omega) \propto q^2$.
Finally, for large $F \gtrsim F_c $, as  in Fig.~\ref{fig4}(c), an anomalous SD-like relaxation appears for all $q<1$. 

\begin{figure}[!tb]
\includegraphics[width=1.0\columnwidth]{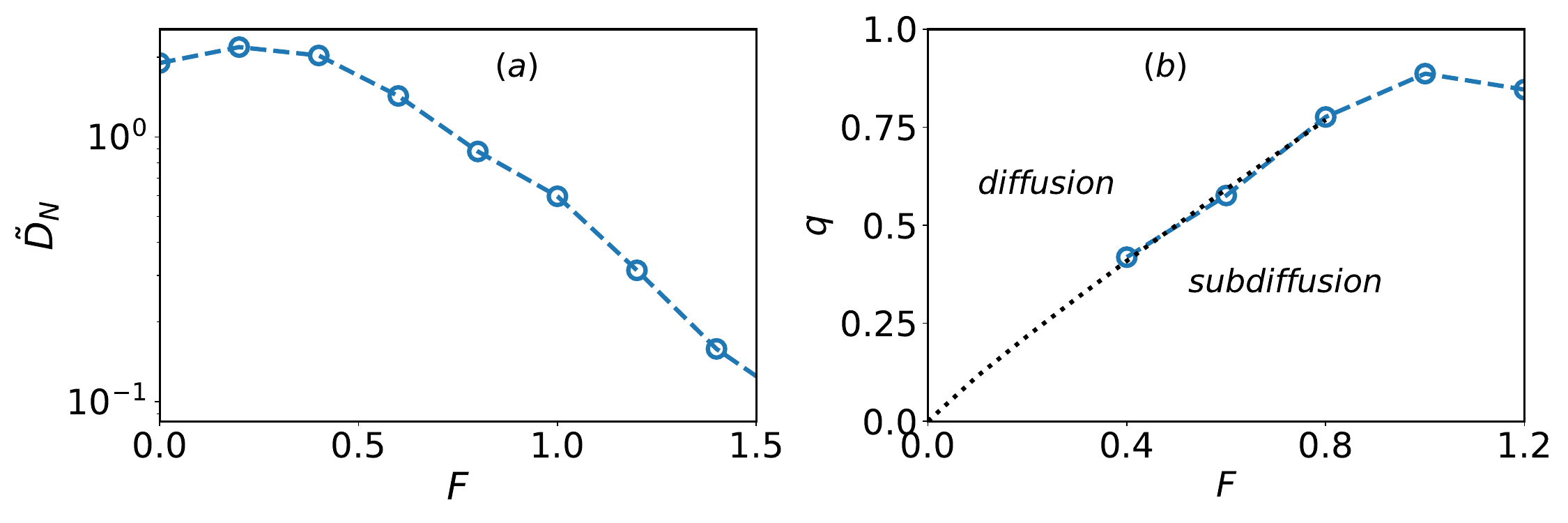}
\caption{ (a) Effective normal diffusion coefficient $\tilde D_N = \Gamma_{q}/q^2$ for $q = \pi/3$
vs. $F$, evaluated  at $q=\pi/3$ via MCLM on $L=24$ sites.  (b) Effective phase diagram $q(F)$ 
with the SD/normal diffusion  crossover at $q^*(F)$, with the dotted line representing the 
qualitative guess for small $F$. }
\label{fig5}
\end{figure}

In Fig.~\ref{fig5}(a) we collect results for an effective normal diffusion coefficient 
$\tilde{D}_N  = {\rm Im} M_q(\omega=0)/q^2$  at largest $q=q_4=\pi/3$ as in Fig~\ref{fig4}.
Taking results for $D_S$ from Fig.~\ref{fig2}(b) and $\tilde{D}_N$ from  Fig.~\ref{fig5}(a) one 
may estimate that the crossover between SD and diffusive relaxations is expected at the wavevector $q^*$ 
fulfilling the relation  $D_S  (q^*)^4=  \tilde{D}_N (q^*)^2$.
Numerical results  from such estimate are shown in Fig.~\ref{fig5}(b), representing a rough phase diagram of
normal-SD transport, relevant at least for moderate $F<F_c$.  
While we expect a continuous vanishing of $q^*$ for $F\to 0$, it is hard to numerically determine the dependence 
$q^*(F \ll 1)$ due to finite-size limitations. Still, the general trend  of $q^*(F)$ 
is well consistent with phenomenological approaches \cite{guardado20,zechmann22}.
We should also note that the explicit MF expression, Eq.~(\ref{mqs}), is restricted  only to the hydrodynamic regime $q \to 0$ 
when $M_q(\omega)$ is small and $\phi_N(\omega \to 0)$ strictly vanishes. Therefore, it cannot be  extended to the discussion of the  normal/SD crossover at larger $q$.

{\it Conclusions.}
In this Letter, we present the analysis of  the Stark models, based on the memory-function
approach to hydrodynamics.  It enables a direct  consideration of the subdiffusion and the corresponding transport  
coefficient $D_S$.  In models with strictly conserved dipole moment $P$,
the derivation yields a subdiffusive relaxation rates, $\Gamma_q = D_S q^4$ and 
an explicit Einstein relation which links $D_S$ with correlations of the uniform ($q=0$)
 dipolar current $J_D$. In the full Stark problem, where 
$P$ is conserved only in the thermodynamic limit, the analogous treatment
is valid only in more restricted  hydrodynamic regime $q \to 0$. However, the  correlations of the
dipolar current are still linked with $D_S$ via the Einstein relation. Moreover,
the obtained numerical results   agree with alternative numerical approaches, 
at least in the range of $F$ where the finite-size limitations allow for reliable numerical studies. 
As a general observation, we note that the coefficient $D_S$ reveals a strong exponential-like reduction
with $F$, which for large $F \gg 1$ can effectively appear as Stark MBL concept 
\cite{schulz19,nieuwenburg19,taylor20,nehra20,zhang21,morong21,doggen21,wang21,gunawardana22,zisling22,wei22},
as well as the Hilbert-space fragmentation \cite{sala20,khemani20,yang20,moudgalya22,moudgalya221,skinner22}.
It should be stressed that the presented Einstein relations are not specific to considered model and even
not to one-dimensional systems. They can be easily generalized to models which are more relevant for experiments
as, e.g.,  the tilted Fermi-Hubbard model. 

\acknowledgements 
We acknowledge the support by Slovenian Research and Innovation Agency (ARIS) via the program 
P1-0044 (P.P., S.N.) and the project J1-2463 (S.N.)
as well as  the National Science Centre, Poland via project 2020/37/B/ST3/00020 (M.M.). 
S. N. acknowledges also the EU support via QuantERA grant
T-NiSQ, and also computing time at the supercomputer Vega at IZUM, Slovenia.
The numerical calculation were partly carried out at the facilities of the Wroclaw Centre for 
Networking and Supercomputing.
Our TEBD code was written
using ITensors Library in Julia \cite{itensor}.
\bibliography{manustarkmf}


\newpage
\phantom{a}
\newpage
\setcounter{figure}{0}
\setcounter{equation}{0}

\renewcommand{\thetable}{S\arabic{table}}
\renewcommand{\thefigure}{S\arabic{figure}}
\renewcommand{\theequation}{S\arabic{equation}}
\renewcommand{\thepage}{S\arabic{page}}

\renewcommand{\thesection}{S\arabic{section}}

\onecolumngrid

\begin{center}

{\large \bf Supplemental Material:\\
Einstein relation for subdiffusive relaxation  in Stark chains}\\

\vspace{0.3cm}

\setcounter{page}{1}

P. Prelov\v{s}ek$^{1}$, S. Nandy$^{1}$, M. Mierzejewski$^{2}$
\\
\ \\
$^1${\it Department of Theoretical Physics, J. Stefan Institute, SI-1000 Ljubljana, Slovenia} \\
$^2${\it Institute of Theoretical Physics, Faculty of Fundamental Problems of Technology, \\ Wroc\l aw University of Science and Technology, 50-370 Wroc\l aw, Poland}\\

\end{center}

\vspace{0.6cm}
In the Supplemental Material we analyze the relaxation of dipole moment and related current correlations,
We present also results for other model parameters, involving the additional numerical approach 
using the  boundary-driven open systems.  \\
\vspace{0.3cm}

\twocolumngrid

\label{pagesupp}

\section{Relaxation of the dipole moment and current correlations} 

Studying the Stark chain in the main text we  argue that for large $L$,  nonzero $F$ and small $\omega$, the dominating contribution to the memory function, Eq. (\ref{mqs}), comes from the dipolar currents, $\phi_{D}(\omega)$, instead of the normal current correlations,  $\phi_{N}(\omega)$. The latter expectation originates from the emergent conservation of the dipole moment in macroscopic systems, $L \to \infty$ \cite{nandy24}.
Below we provide numerical results which support this claim. However, we note that in a {\em finite} system one gets $\phi_{N}(\omega= 0)=0$ just due to open boundary conditions (OBC) simply because a finite system with OBC cannot host a steady current. In order to disentangle the influence of OBC in a finite system from the conservation of the dipole moment we study the relaxation of the   dipole moment \mbox{$P = \sum_l (l-L/2)  n_l$}. To this end we calculate the correlation function 
\begin{equation} 
\phi_P(\omega) =(P |({\cal L} -\omega)^{-1} | P)= \frac{ - \chi^0_P}{\omega +N(\omega)}. \label{phip}
\end{equation}
In equation (\ref{phip}) we  introduced the  corresponding MF
\begin{equation} 
N(\omega)= \frac{L}{\chi^0_P} \tilde \phi_N(\omega), \label{phip1}
\end{equation}
where $\chi^0_P = \langle P^2 \rangle $ and formally MF can be expressed as \cite{forster75},
\begin{equation} 
\tilde \phi_N(\omega)=({\cal L} P |({\cal \tilde{L}} -\omega)^{-1} | {\cal L} P)/L=(J_N |({\cal \tilde{L}} -\omega)^{-1} | J_N)/L,  \label{phint}
\end{equation}
where we used the identity ${\cal L} P= -i J_N$.
In contrast to Eq.~(\ref{mf}) in the main text that is valid in the hydrodynamic (perturbative) regime, there is no small parameter
in Eq.~(\ref{phint}) thus $\cal \tilde{L}$  contains additional projectors \cite{forster75},
 \begin{equation} 
{\cal \tilde{L}}={\cal Q L Q} , \quad {\cal Q}=1-{\cal P}, \quad {\cal P}=\frac{| P)(P|}{(P|P)}.
\end{equation}
For this reason in a finite system $\tilde \phi_N(\omega) $ in Eq. (\ref{phint})  is not equal to 
$\phi_N(\omega) = (  J_N | ({\cal L}  - \omega)^{-1} |  J_N)/L$ but both quantities should
merge for $L \to \infty $ due to the conservation of $P$.

\begin{figure}[!tb]
\includegraphics[width=0.8\columnwidth]{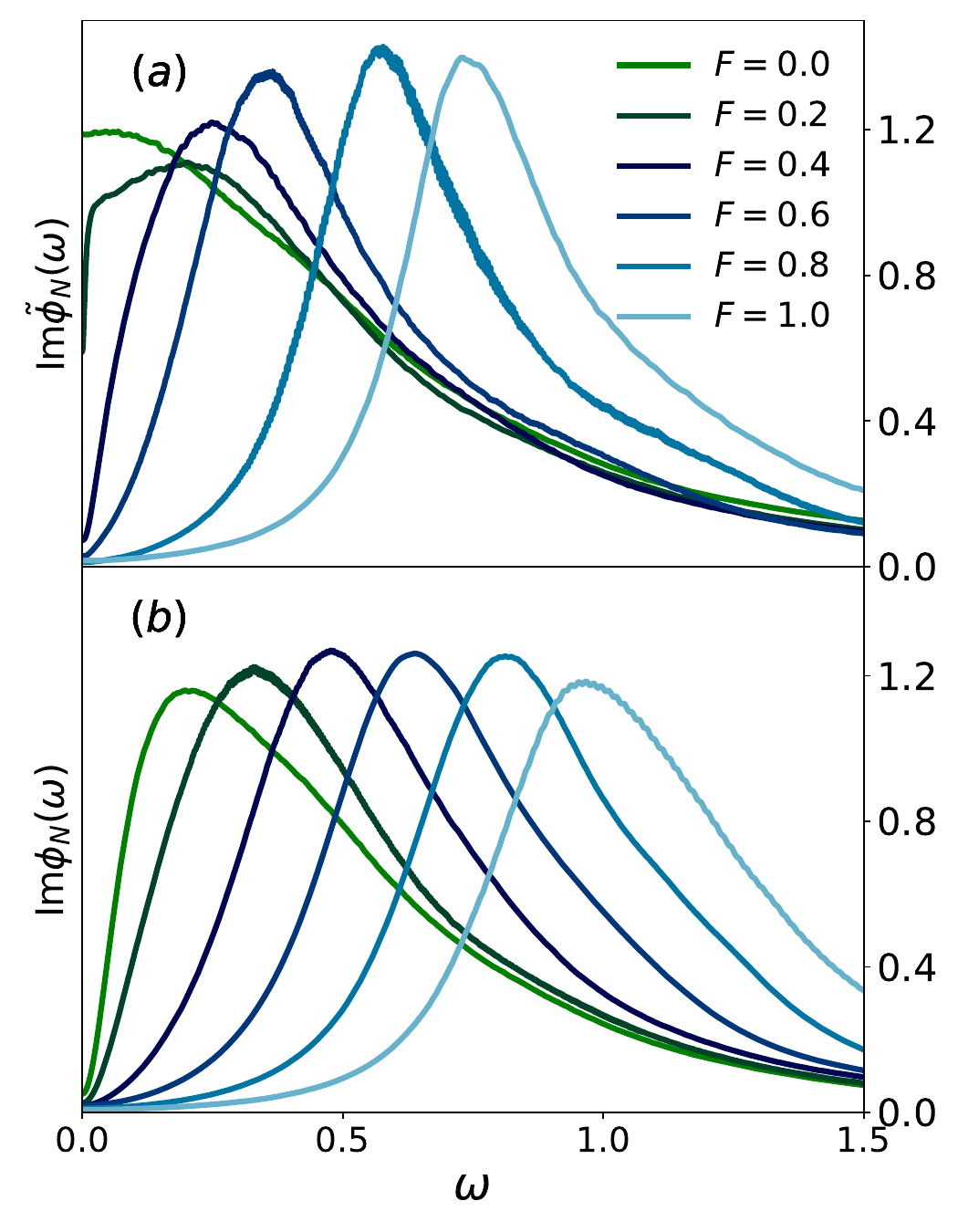}
\caption{(a) The current correlation function $\mathrm{Im} \tilde \phi_N(\omega)$, as extracted form
dipole-moment correlations $\phi_P(\omega)$, calculated for Stark model using
MCLM on $L=28$ chain, for different fields $F = 0 - 1.0$, (b) directly evaluated 
current correlations $\mathrm{Im} \phi_N(\omega)$ in the same system. 
}
\label{figS1}
\end{figure}

Due to the presence of projectors, it is hard to determine  $\tilde \phi_N(\omega) $ from Eq. (\ref{phint}). 
However, this quantity can be obtained also from the correlations of the dipole moment via Eqs. (\ref{phip}) and  (\ref{phip1}).
In order to confirm the relation between  $\tilde \phi_N(\omega) $ and $\phi_N(\omega)$ 
and related conjectures, we numerically investigate the Stark model 
with the same parameters $V=V'=1$ for different fields $F \le 1$.
On one hand, using MCLM on the $L=28$ system with OBC,  we evaluate $\phi_P(\omega)$ and extract the
corresponding $\mathrm{Im} \tilde \phi_N(\omega)$. On the other hand,
 we calculate within the same system directly $\mathrm{Im} \phi_N(\omega)$. Numerical results presented in 
 Figs.~\ref{figS1}(a) and  \ref{figS1}(b) confirm that  at $F > F^* \sim 0.4$ (for considered $L=28$) both approaches
indeed yield (at least qualitatively)  $\tilde \phi_N(\omega) \sim \phi_N(\omega)$, up to finite-size effects. 
As expected, the disagreement occurs at $F \sim 0$ since $\phi_N(\omega \to 0)$ vanishes due to OBC, while 
$\tilde \phi_N(\omega \to 0)$ does not.  Due to the same argument the qualitative discrepancy in the range
 $ F<F^*(L)$ can be attributed to  finite-size limitations.
 
The main conclusion is  that $P$ behaves as a conserved quantity
and, at the same time, $\mathrm{Im} \phi_N(\omega \to 0) \to 0$, so that
in Eq.~(\ref{mqs}) in the main text one can neglect the latter contribution in the hydrodynamic regime $ q \to 0$. 
 Consequently, the relaxation is then dominated by the SD  contribution,
 i.e., $M_q(\omega \sim 0) \sim q^4 \phi_D(\omega \to 0)/\chi^0_0$.
 Based on results in Fig.~\ref{figS1} this conclusion holds true  
  at least for $F>F^*\sim 0.4$, whereby we expect that $F^*$ vanishes in the thermodynamic limit.
 
\section{Subdiffusion coefficient from boundary driven open systems}

As an additional technique to compute the subdiffusion constant, we study an open chain described by Eq.~(\ref{stark}), driven via boundary Lindblads operators inducing a weak particle current by creating a bias $\mu$ at the edges of the system. The master equation governing the time-evolution of the density matrix is given by 
\begin{eqnarray}
\partial_{t}\rho = -i[H, \rho] + \hat{\mathcal{D}}\rho
\label{lindblad_equation}
\end{eqnarray}
where $\hat{\mathcal{D}}$ denotes the dissipator operator written in terms of the Lindblads as $\hat{\mathcal{D}}\rho=\sum_{k}L_{k}\rho L_{k}^{\dagger} - \frac{1}{2}\{L_{k}^{\dagger}L_{k}, \rho\}$. The details of the Lindblads employed here are discussed in \cite{nandy23,nandy24, PrelovsekPRB2022}. To evolve the density
matrix towards the steady state $\rho_{ss}$, we use the time-evolving
block decimation (TEBD) for vectorized density matrices. In particular, we use the fourth-order TEBD with a time step $dt = 0.2$, bond dimension $\chi \sim 140$, and weak bias $\mu = 0.01$.

The method allows us to establish a nonequilibrium steady state (NESS) for which the normalized current $I/\mu$ and the spatial profile of particles $\langle \tilde{n}_{l} \rangle$ can be easily computed. One remarkable advantage of this technique is that the NESS profile carries the signature of the nature of particle transport \cite{nandy24}. In particular, for system with both particle-number and dipole-moment conservation, the hydrodynamic equation of motion $\partial_{t}n + D\partial^{4}_{x}n=0$ implies the NESS particle profile of the form $\langle \tilde{n}_{l} \rangle = ax + bx^{3}, x=2\ell/L-1$. 

\begin{figure}[!tb]
\includegraphics[width=0.8\columnwidth]{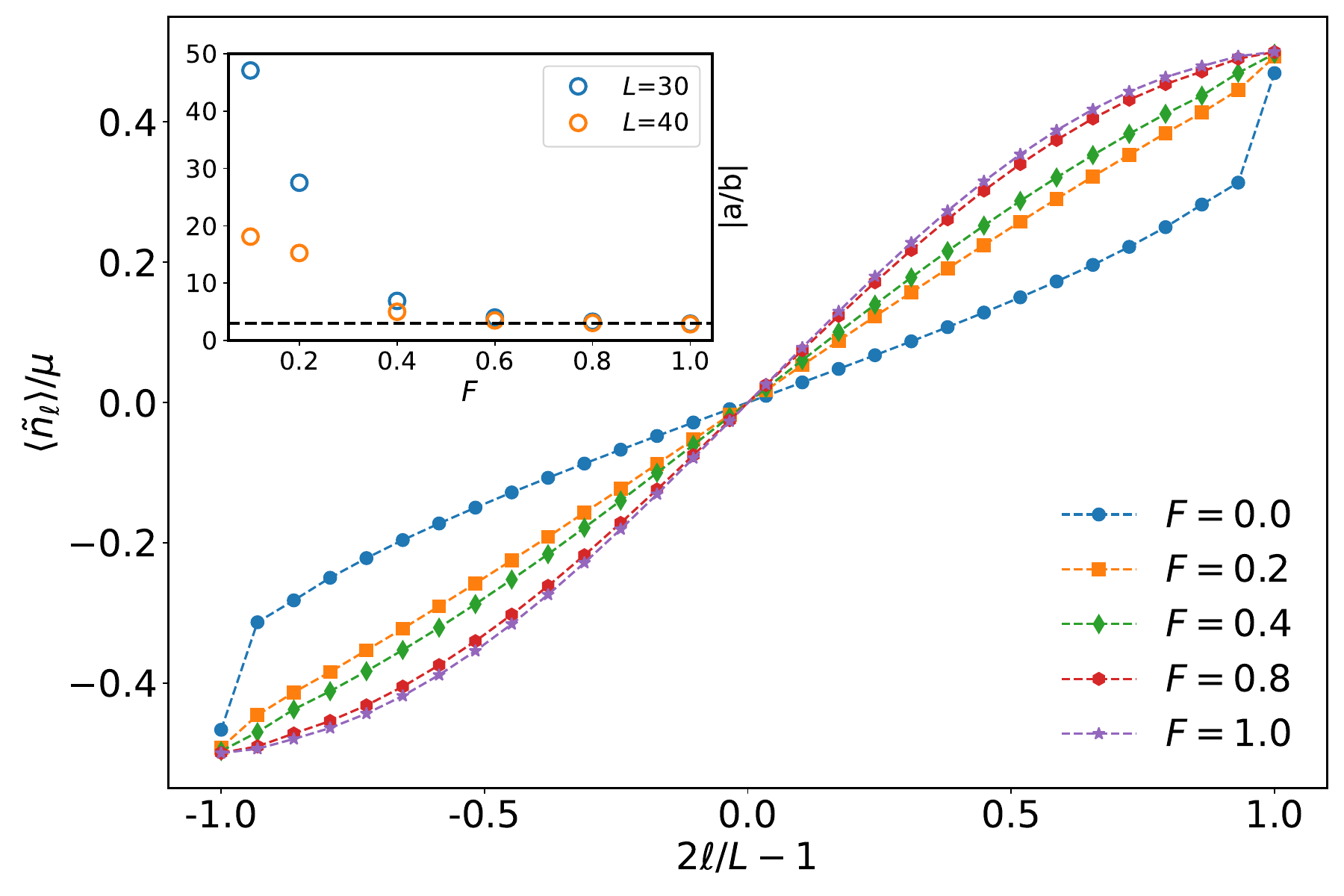}
\caption{Steady-state density profiles $\langle \tilde n_l \rangle/\mu$  obtained by the boundary-driven open-system 
approach, obtained for different fields $F$ on $L=40$ system. Inset shows the ratio of profile coefficients $|a/b|$ vs.  
$F$ for two system sizes $L$.}
\label{figS2}
\end{figure}

We further present and analyse the NESS results for the case $V=2, V^{\prime}=0$, which is at $F=0$ 
equivalent to the isotropic Heisenberg  model with the corresponding particle superdiffusion \cite{nandy23},  while for $F >0$ the SD
is expected with  results similar as presented in the main text for $V = V'=1 $.
Fig. \ref{figS2} clearly demonstrates how the normalized particle profile changes with increasing $F$. The profile at $F=0$ is indeed a signature of particle superdiffusion where $\langle \tilde{n}_{l} \rangle \sim \arcsin(x)$ (see \cite{nandy23} for details). 
As evident in Fig. \ref{figS2}, increasing $F > 0$ leads to qualitative change of the profile. Whereas it is close to linear 
(corresponding to normal diffusion) for small finite 
$F \lesssim 0.4$, eventually the profile becomes cubic for larger $F \gtrsim 0.6$, being a direct signature of the  crossover to 
SD transport. Here, we refer to reference \cite{nandy24}, for further details regarding the fitting of coefficients $a, b$. 
Finally, for large $F$ with the SD transport we establish the ratio $|a/b| \sim 3$, as the consequence of vanishing derivative 
$\delta_x(ax+bx^3)|_{x = \pm 1}=0$ at the edges, being evident in the main panel of Fig. \ref{figS3}. The inset shows that indeed the ratio $|a/b| \to 3$ with increasing $F$ and also with increasing $L$. We argue that the edge flattness carries the signature of the fact that in 
SD systems the normal current becomes zero (or negligibly small) for $F>0$ in the $L \to \infty$ limit. Using the relation $|a/b|=3$,  the SD coefficient can be then expressed directly with the obtained NESS current as  $D_S = IL^{3}/(12\mu)$. 

\section{Results for other model parameters}

\begin{figure}[!tb]
\includegraphics[width=0.9\columnwidth]{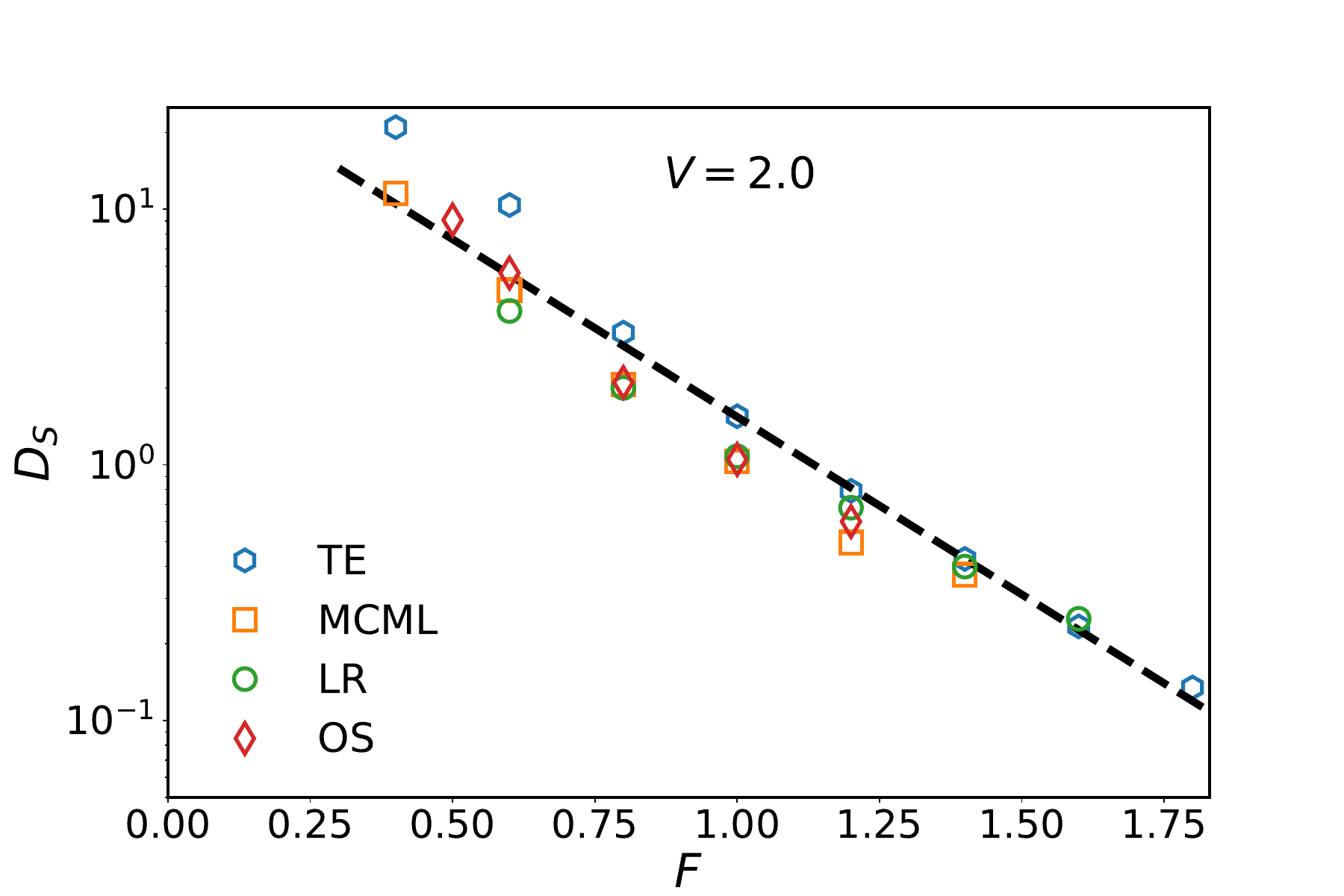}
\caption{Subdiffusion coefficient $D_S(F)$, as calculated:  
from rates $\Gamma_{q_1}$, obtained via time-evolution (TE) and via MCLM on $L=24$ chain, 
and directly from Einstein relation (ER) and $\phi_D(\omega)$ on $L=28$ chain, as well 
via the boundary-driven open systems (OS) on $L \leq 50$ chains. }
\label{figS3}
\end{figure}

We present in Fig.~\ref{figS3} the summary of results for the SD coefficient $D_S$,
as obtained by different numerical approaches for parameters $V =2, V'=0$, corresponding to
the isotropic Heisenberg model subject to constant field (magnetic field gradient). 
Besides the methods presented and used in the main text, i.e., the time-evolution (TE)
and MCLM where $D_S$ is extracted from relaxation rates $D_S = \Gamma_{q_1}/q_1^4$
for the smallest $q_1=2 \pi/L$ at $L=24$, and the direct evaluation using the Einstein relation (ER)
on $L=28$ chain, we include now also the NESS results for the boundary-driven systems 
with $L \lesssim 50$. Results are presented for fields $F > F^*(L) \sim 0.4$
where the SD transport is dominant for numerically accessible $L$. Employed methods
have also different numerical limitations at large $F > 1$. While the NESS approach is limited by smallness of  the
current $I$, the MCLM and Einstein relation are limited by the frequency resolution and the influence of OBC, 
so the time-evolution approach has the advantage of smallest reachable $\Gamma_q$ and consequently 
of largest reachable $F \lesssim 2$ due to PBC in this case.  The overall conclusion following from Fig.~\ref{figS3} 
is that results by different methods are even quantitatively well consistent, revealing 
again the strong/exponential dependence of $D_S$ on $F$.  On the other hand, by comparing with 
Fig.~\ref{fig2} (b) in the main text, we also notice that $D_S$ only weakly depends on the interaction parameters $V$ and  $V'$.  
For stronger interactions we observe an  increase $D_S$  at larger $F$ since  the interaction suppresses localization, while 
at weaker $F$  the trend might be opposite, but further studies are needed to establish such relations.

\end{document}